\documentclass[aps,prl,reprint,showpacs,floatfix]{revtex4-1}

\usepackage{amsmath,amssymb}
\usepackage{graphicx}

\usepackage{bm}

\newcommand{\figurewidth}{0.42\textwidth}

\newcommand{\epslj}{\varepsilon_{\mathrm{LJ}}}
\newcommand{\siglj}{\sigma_{\mathrm{LJ}}}

\begin{document}

\title{Dielectric effects in the self-assembly of binary colloidal
  aggregates}

\author{Kipton Barros}
\affiliation{Department of Materials Science \& Engineering and
  Department of Engineering Sciences \& Applied Mathematics,
  Northwestern University, Evanston, Illinois 60208, U.S.A.}

\author{Erik Luijten}
\email[Corresponding author: ]{luijten@northwestern.edu}
\affiliation{Department of Materials Science \& Engineering and
  Department of Engineering Sciences \& Applied Mathematics,
  Northwestern University, Evanston, Illinois 60208, U.S.A.}

\begin{abstract}
  Electrostatic interactions play an important role in numerous
  self-assembly phenomena, including colloidal aggregation. Although
  colloids typically have a dielectric constant that differs from the
  surrounding solvent, the effective interactions that arise from
  inhomogeneous polarization charge distributions are generally
  neglected in theoretical and computational studies. We introduce an
  efficient technique to resolve polarization charges in dynamical
  dielectric geometries, and demonstrate that dielectric effects
  \emph{qualitatively} alter the predicted self-assembled structures,
  with surprising colloidal strings arising from many-body effects.
\end{abstract}

\date{March 19, 2013 --- Final version June 5, 2014}

\pacs{77.84.Nh, 82.70.Dd, 61.20.Ja, 77.22.Ej}

\maketitle

Colloids are ubiquitous in systems of physical, chemical, and biological
interest.  In suspension, dissociation of surface groups frequently
causes these particles to carry an electrical charge, resulting in
electrostatic interactions that play an important role in colloidal
stability, aggregation, and
self-assembly~\cite{russel89,hunter01,leunissen05}.  Far less is known
about the effect of induced polarization charges.  Although molecular
dynamics (MD) and Monte Carlo simulations of charged colloids are now
commonplace, they rarely take into account dielectric effects and
instead treat the dielectric constant as spatially uniform.  This is
particularly striking in view of the large dielectric contrast between
typical colloids and an aqueous solution (e.g., $\kappa \approx 2.5$ for
polystyrene \emph{vs.}\ $\kappa \approx 80$ for water at $293$~K), which
induces significant polarization charges at the colloidal surface.
Densely packed and anisotropic arrangements of dielectric objects make
this approximation even less justified.  Thus, there is a pressing need
to assess the role of dielectric effects in self-assembly phenomena.

Proper treatment of dielectric effects has been limited by computational
complexity.  Only the simplest dielectric geometries permit analytical
solution.  For an interacting system of dielectric spheres, a series
expansion has been derived~\cite{doerr06}, but this still requires
expensive numerical evaluation. A more general approach is to
numerically solve the induced bound charge self-consistently over
discretized dielectric
interfaces~\cite{levitt78,hoyles98,allen01,boda04,tyagi10,jadhao12}.
This approach does not constrain the geometry, but has not yet been
efficient enough to allow simulation of dynamical dielectric objects,
such as mobile colloids.  Indeed, existing work has largely treated the
dielectric geometry as static, focusing on ion distributions in
planar~\cite{croxton81,torrie82,kjellander85} or
spherical~\cite{messina02a,rescic08} geometries.

In this Letter, we address this situation by presenting the first study
of a dielectric system with a \emph{fully dynamic} geometry, exploring
the effect of polarization charges that respond to and influence the
motion of charged colloids.  Using an optimized simulation
method~\cite{barros14a} we explicitly demonstrate that dielectric
interactions can \emph{qualitatively} alter self-assembly in a
prototypical size-asymmetric binary mixture of charged colloids in
solution. In particular, polarization charge that binds a colloid pair
can also effect repulsive three-body interactions, giving rise to
string-like colloidal chains.

\begin{figure}
\includegraphics[width=\figurewidth]{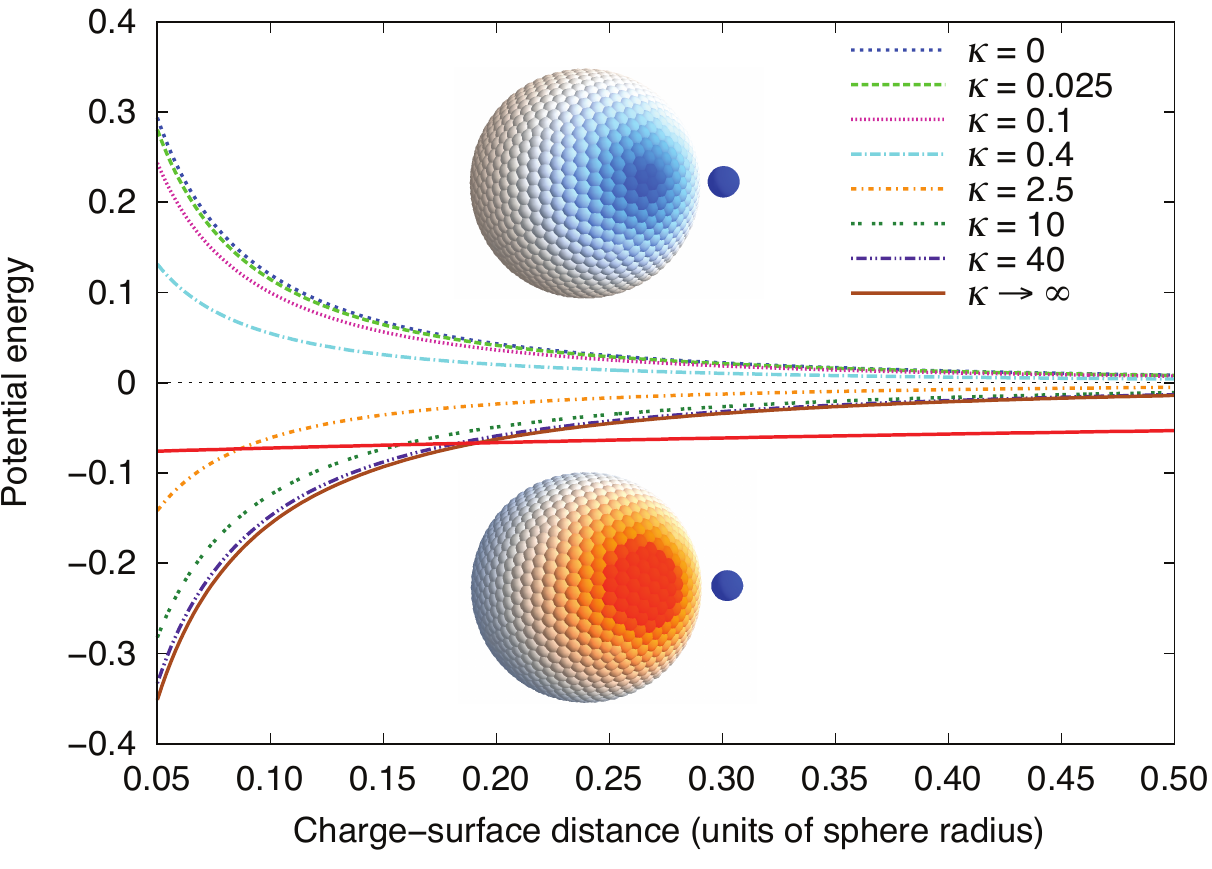}
\caption{Electrostatic energy (in units of $q^2 / (\varepsilon_0
  \kappa_{\mathrm{m}}R)$) of a \emph{neutral} sphere of radius~$R$
  and dielectric constant $\kappa_{\mathrm{obj}}$ and a negative
  point charge~$q$ embedded in a medium of dielectric
  constant~$\kappa_{\mathrm{m}}$, as a function of ion--surface
  separation, for different values of the reduced dielectric
  constant~$\tilde \kappa = \kappa_{\mathrm{obj}} /
  \kappa_{\mathrm{m}}$.  The induced bound charges repel the point
  charge for $\tilde \kappa < 1$ (top inset; color coding represents calculated
  polarization charge density), whereas for $\tilde \kappa > 1$ the induced
  charges are attractive (bottom inset).  The near-horizontal solid line
  indicates the pure Coulomb interaction for a reference system of two
  oppositely charged nondielectric spheres.}
\label{fig:sphereenergy}
\end{figure}

To gain insight in the role of dielectric mismatch between colloidal
particles and the surrounding solvent, we briefly review systems of
linear dielectrics, starting from the electrostatic (free)
energy~\cite{jackson99},
\begin{equation}
  U = \frac{1}{2} \int \rho_f(\mathbf{r}) \psi(\mathbf{r}) 
    \,\mathrm{d}\mathbf{r} \;,
  \label{eq:energy}
\end{equation}
where $\rho_f(\mathbf{r})$ is the free charge density and the potential
$\psi(\mathbf{r})$ is defined through Poisson's equation,
\begin{equation}
  \nabla \cdot \left[\kappa(\mathbf{r}) \nabla \psi(\mathbf{r})\right]
  = -\rho_f(\mathbf{r}) / \varepsilon_0 \;,
  \label{eq:constitutive} 
\end{equation}
with $\kappa(\mathbf{r})$ the material-specific and spatially varying
dielectric constant and $\varepsilon_0$ the vacuum permittivity.  If we
scale $\kappa \rightarrow \gamma \kappa$ and $\rho_f \rightarrow \alpha
\rho_f$ ($\alpha, \gamma > 1$), the energy scales as $U \rightarrow
(\alpha^2/\gamma) U$, so that the behavior of a system is invariant if
$\gamma = \alpha^2$.  Here, we are interested in dispersions of
colloidal particles with $\kappa = \kappa_{\mathrm{obj}}$ in a medium
(solvent) with $\kappa = \kappa_{\mathrm{m}}$. Such a system is
mathematically equivalent to colloids with \emph{reduced} dielectric
constant $\tilde \kappa = \kappa_{\mathrm{obj}} / \kappa_{\mathrm{m}}$
and scaled free charge density $\tilde{\rho}_f = \rho_f /
\sqrt{\kappa_{\mathrm{m}}}$ dispersed in a nonpolarizable solvent.
Thus, without loss of generality, we vary only $\tilde \kappa$ in our
calculations. To illustrate the role of this reduced dielectric
constant, we consider the electrostatic energy of a neutral sphere of
dielectric constant~$\kappa_{\mathrm{obj}}$ and radius~$R$ and a point
charge~$q$ at a distance~$d>0$ from its surface~\cite{jackson99},
\begin{equation}
  U_{\mathrm{sphere}} = \frac{q^2}{8\pi \varepsilon_0
    \kappa_{\mathrm{m}}R} \sum_{n = 0}^{\infty}
  \frac{(1-\tilde \kappa) n}{(1+\tilde \kappa) n + 1}
  \frac{1}{(1 + d/R)^{2(n + 1)}} \;.
\label{eq:sphereeq}
\end{equation}
Depending on $\tilde \kappa$, $U_{\mathrm{sphere}}$
(Fig.~\ref{fig:sphereenergy}) ranges from attractive to
repulsive~\cite{israel2011}.  If $\tilde \kappa > 1$, the induced
surface bound charge closest to the point charge has the opposite sign
as the point charge, and the dielectric effects are attractive (bottom
inset). Conversely, if $\tilde \kappa < 1$, the induced bound charge
leads to repulsive dielectric effects (top inset).  The two limits
$\tilde \kappa =\{0, \infty\}$ correspond to a conducting solvent and a
conducting sphere, respectively, but it is noteworthy that dielectric
effects saturate well before either limit is reached.  The asymmetry
between $\tilde \kappa > 1$ and $\tilde \kappa < 1$ provides the
starting point for exploring the effect of dielectric mismatch on
colloidal aggregation.  However, as we discuss below, physically rich
behavior arises from the many-body interactions and the associated
constraint that the net polarization charge on each colloid is fixed.

In our numerical treatment, we solve for the bound-charge density
$\rho_b(\mathbf{r}) = -\nabla \cdot
\mathbf{P}(\mathbf{r})$. Substitution of the polarization field
$\mathbf{P}(\mathbf{r}) = \varepsilon_0 (\kappa(\mathbf{r}) - 1)
\mathbf{E}(\mathbf{r})$ and the electric field $\mathbf{E}(\mathbf{r}) =
- \nabla \psi(\mathbf{r})$ yields $\rho_b(\mathbf{r}) / \varepsilon_0 =
\nabla \cdot [(\kappa(\mathbf{r}) - 1) \nabla
\psi(\mathbf{r})]$. Comparison with Eq.~\eqref{eq:constitutive}
reproduces the well-known result
\begin{equation}
  \nabla^2 \psi(\mathbf{r})
  = -[\rho_f(\mathbf{r}) + \rho_b(\mathbf{r})] / \varepsilon_0 \;.
  \label{eq:gauss} 
\end{equation}
We define $\mathcal{G}$ to represent the inverse of the operator
$-\nabla^2$. Its explicit action is $\mathcal{G} \rho(\mathbf{r}) =
\frac{1}{4 \pi} \int \frac{\rho ( \mathbf{r}')}{| \mathbf{r} -
  \mathbf{r}' |} \mathrm{d}\mathbf{r}'$.  Equations
\eqref{eq:constitutive} and~\eqref{eq:gauss} combined relate the free
and bound charge,
\begin{equation}
  \nabla
  \cdot [\kappa(\mathbf{r}) \nabla \mathcal{G} (\rho_b(\mathbf{r}) +
  \rho_f(\mathbf{r}))] = -\rho_f(\mathbf{r}) \;,
  \label{eq:Poisson-pol}
\end{equation}
which can be rewritten as~\cite{hoshi87}
\begin{equation}
  \mathcal{A}(\mathbf{r}) \rho_b(\mathbf{r}) = b(\mathbf{r}) \;,
  \label{eq:operatoreq} 
\end{equation}
where $\mathcal{A}(\mathbf{r})$ represents the linear operator
\begin{eqnarray}
  \mathcal{A}(\mathbf{r}) 
  & = & - \nabla \cdot \kappa(\mathbf{r}) \nabla \mathcal{G}
  = \kappa(\mathbf{r}) 
  - (\nabla \kappa(\mathbf{r})) \cdot \nabla \mathcal{G}  
  \label{eq:adef}\\
  b(\mathbf{r}) & = & (1 - \mathcal{A}(\mathbf{r})) \rho_f(\mathbf{r}) \;.
  \label{eq:bdef}
\end{eqnarray}
Equation~\eqref{eq:operatoreq} will be solved for $\rho_b(\mathbf{r})$,
from which the potential, $\psi(\mathbf{r}) = \mathcal{G}
(\rho_f(\mathbf{r}) + \rho_b(\mathbf{r})) / \varepsilon_0$, and other
derived quantities follow.  Equation~\eqref{eq:Poisson-pol} implies that
the net charge in a compact region $\Omega$ with a \emph{uniform}
dielectric constant $\kappa$ on its boundary
is~\cite{barros14a}
\begin{equation}
  \int_{\Omega} [\rho_f(\mathbf{r}) + \rho_b(\mathbf{r})]
  \,\mathrm{d}\mathbf{r} =
  \kappa^{-1}  \int_{\Omega} \rho_f(\mathbf{r})
  \,\mathrm{d}\mathbf{r} \;.
  \label{eq:netcharge} 
\end{equation}
As a consequence, the bound charge in regions of uniform~$\kappa$ is
simply $\rho_b(\mathbf{r}) = (\kappa^{-1} - 1) \rho_f(\mathbf{r})$. The
difficult (and generally ignored) task is to calculate
$\rho_b(\mathbf{r})$ when $\nabla \kappa(\mathbf{r}) \neq 0$.  We
consider systems with sharp material interfaces, where the bound-charge
density has to be calculated at the interface rather than in the entire
volume---a considerable numerical simplification~\cite{allen01}. The
strategy is to solve Eq.~\eqref{eq:operatoreq} as a discretized matrix
equation for the \emph{surface} charge density~$\sigma(\mathbf{r})$
\cite{hoshi87,boda04},
\begin{equation}
  \mathcal{A}_{ij} \sigma_j = b_i \;.
  \label{eq:matrixeq} 
\end{equation}
This matrix equation has the same mathematical content as previous
discretizations~\cite{levitt78,hoyles98,allen01}.

However, in a \emph{dynamical} situation, where dielectric objects move,
$\mathcal{A}_{ij}$ is evolving via its dependence on the dielectric
geometry~$\kappa(\mathbf{r})$. At each time step, the explicit
construction of $\mathcal{A}_{ij}^{-1}$ would require $\mathcal{O}(N^3)$
operations, where $N$ is the number of discretized surface patches.
Since this is prohibitively expensive, we instead opt to solve
Eq.~\eqref{eq:matrixeq} for $\sigma_j$ using an iterative
method~\cite{boda04}.  As shown in Ref.~\cite{barros14a}, iterative
methods~\cite{saad86,vandervorst92}
are desirable for two reasons: (i)~explicit construction of the matrices
$\mathcal{A}_{ij}$ or $\mathcal{A}_{ij}^{-1}$ is not required and the
cost of each iteration scales as the cost of the matrix--vector product
$\mathcal{A}_{ij} x_j$; (ii)~convergence requires only a few iterations
because the eigenvalues of $\mathcal{A}_{ij}$ have a favorable
structure. The only expensive, nonlocal piece of $\mathcal{A}_{ij} x_j$
(cf.\ Eq.~\eqref{eq:adef}) is the calculation of
$\nabla\mathcal{G}x$---essentially finding $\mathbf{E}(\mathbf{r})$ for
a given charge distribution~$x(\mathbf{r})$.  With an efficient Ewald
solver one can numerically evaluate $\nabla\mathcal{G} x$ with
$\mathcal{O}(N)$~\cite{greengard97} or $\mathcal{O}(N \ln
N)$~\cite{sagui01} operations~\cite{tyagi10}.  The number of iterations
required to solve Eq.~\eqref{eq:matrixeq} at fixed numerical accuracy is
bounded by $\log | \lambda_{\max} / \lambda_{\min} |$, the ratio of the
largest and smallest eigenvalues of $\mathcal{A}_{ij}$. Since
$\mathcal{A}_{ij}$, although neither symmetric nor normal, is
diagonalizable with positive real eigenvalues that are bound by the
extremal dielectric constants that occur in the system, $\kappa_{\min}
\leqslant \lambda \leqslant \kappa_{\max}$~\cite{barros14a}, the number
of iterations scales at most as $\log [\kappa_{\max} / \kappa_{\min}]$,
where typically $\kappa_{\max} / \kappa_{\min} \lesssim 25$.
Furthermore, the worst-case convergence rates occur only in geometries
with extreme aspect ratios, such as the infinite dielectric slab or
cylinder.  Employing GMRES, which requires only one matrix--vector
product ($\mathcal{A} x$) per iteration and minimizes the residual in
each iteration, we typically reach convergence ($10^{-4}$ relative error
in the electrostatic energy) within five iterations for a system of
spherical objects---achieving a far higher efficiency than prior
approaches. For comparison, the iterative methods in
Ref.~\cite{levitt78,hoyles98,allen01,tyagi10} essentially reduce to
Richardson iteration, which converges more slowly and requires manual
tuning of a relaxation parameter; if this parameter is not properly
chosen, the method may even diverge.
To make progress in simulating mobile dielectric objects, several
additional steps are needed to ensure efficiency and accuracy. In each
iteration we constrain the net charge on each object to its correct
value via Eq.~\eqref{eq:netcharge}, thus eliminating a slow relaxation
mode of the iterative solver and simultaneously improving the precision
of the polarization charges. Furthermore, we replace the internal (free)
charge~$q$ inside each object with a distribution of the
charge~$q/\kappa_{\mathrm{m}}$ that generates the same potential outside
the object.

Lastly, in this first numerical study of mobile dielectrics we must
address the electrostatic \emph{force on a dielectric object}. This includes
forces between induced and free charges as well as forces between
polarization charges induced on different objects, and the resulting
torques that may arise.  A calculation of this force from first
principles requires taking the derivative of the energy
Eq.~\eqref{eq:energy} with respect to object
position~\cite{barros14a}.
If the free charge is rigidly fixed to the dielectric object this yields
\begin{equation}
  \mathbf{F} = - \nabla U 
  = \kappa_{\mathrm{m}} \int_{\Omega} 
  \mathbf{E}(\mathbf{r}) (\rho_f(\mathbf{r}) + \rho_b(\mathbf{r}))
  \, \mathrm{d}\mathbf{r} \;,
  \label{eq:force} 
\end{equation}
where $\Omega$ extends over the object. If the dielectric constant of
the object matches the solvent, the integrand reduces to the electric
component of the Lorentz force density, $\mathbf{f}(\mathbf{r}) =
\kappa_{\mathrm{m}} \mathbf{E}(\mathbf{r}) ( \rho_f(\mathbf{r}) +
\rho_b(\mathbf{r}) ) = \mathbf{E}(\mathbf{r}) \rho_f(\mathbf{r})$.  A
simple physical argument supporting Eq.~\eqref{eq:force} follows from
the principle of effective moments~\cite{pohl51,jones95} where one
replaces the dielectric object with a virtual distribution of free
charge $\rho_v(\mathbf{r})$ that preserves the potential
$\psi(\mathbf{r})$ external to the domain $\Omega$ of the object.  A
correct choice is indeed $\rho_v(\mathbf{r}) = \kappa_{\mathrm{m}}
(\rho_f(\mathbf{r}) + \rho_b(\mathbf{r}))$.
The torque follows naturally from the force density.

The strategy outlined here now allows us to investigate a prototypical
system of electrostatic self-assembly, namely a size-asymmetric binary
mixture of charged spherical colloids.  Such suspensions occur in a
variety of
contexts~\cite{romero-enrique00,liu04b,leunissen05,vantomme08}. It is
important to note that we choose this salt-free model system to
highlight dielectric effects. Experimental realizations generally
contain salt, which screens the electrostatic interactions and thereby
diminishes the role of polarization. We also disregard van der Waals
interactions, notably Debye induction interactions (of similar origin as
the induced interactions considered here, but far
weaker~\cite{israel2011}) and London dispersion forces (which display
many-body effects as well~\cite{kim06,cole10}). Our model constrains the
colloids to have constant charge. In reality, charge regulation, in
which the colloidal surfaces have a dynamic ionization state, provides a
more accurate description than either constant-charge or
constant-potential boundary conditions~\cite{ninham71,chan83,popa10},
but its incorporation in a many-particle simulation would extend the
computational complexity even further; furthermore, the present model
offers the advantage of isolating the dielectric effects, permitting a
\emph{quantitative} assessment of their relevance compared to the
nonpolarizable models widely employed in colloidal self-assembly.

The solvent (dielectric constant~$\kappa_\mathrm{m}$) contains an equal
mixture of small colloids (diameter~$\siglj$, free charge~$-q$,
dielectric constant~$\kappa_{\mathrm{small}}$) and large colloids
(diameter~$7\siglj$, $+q$, $\kappa_\mathrm{large}$). The bound charge on
the small colloids is distributed uniformly, and its fluctuations are
assumed to be small.  Indeed, we have verified that full treatment of
these fluctuations would lead to corrections $\lesssim 1\%$ in the
pair energy and $\sim 5\%$ in the pairwise forces~\cite{barros14a}.  For a
large--small pair at separation $r \gg R = 3.5 \siglj$, the induced
interactions decay as $r^{-4}$, much faster than the direct Coulombic
interactions.  However, as Fig.~\ref{fig:sphereenergy} shows, at small
separations $r \approx R$ dielectric interactions become important,
reaching a magnitude comparable to the Coulombic interactions at contact
($d = 0.5 \siglj / 3.5 \siglj \approx 0.14$) for $\tilde \kappa \gg 1$
or $\tilde \kappa \ll 1$.

\begin{figure}
\includegraphics[scale=1]{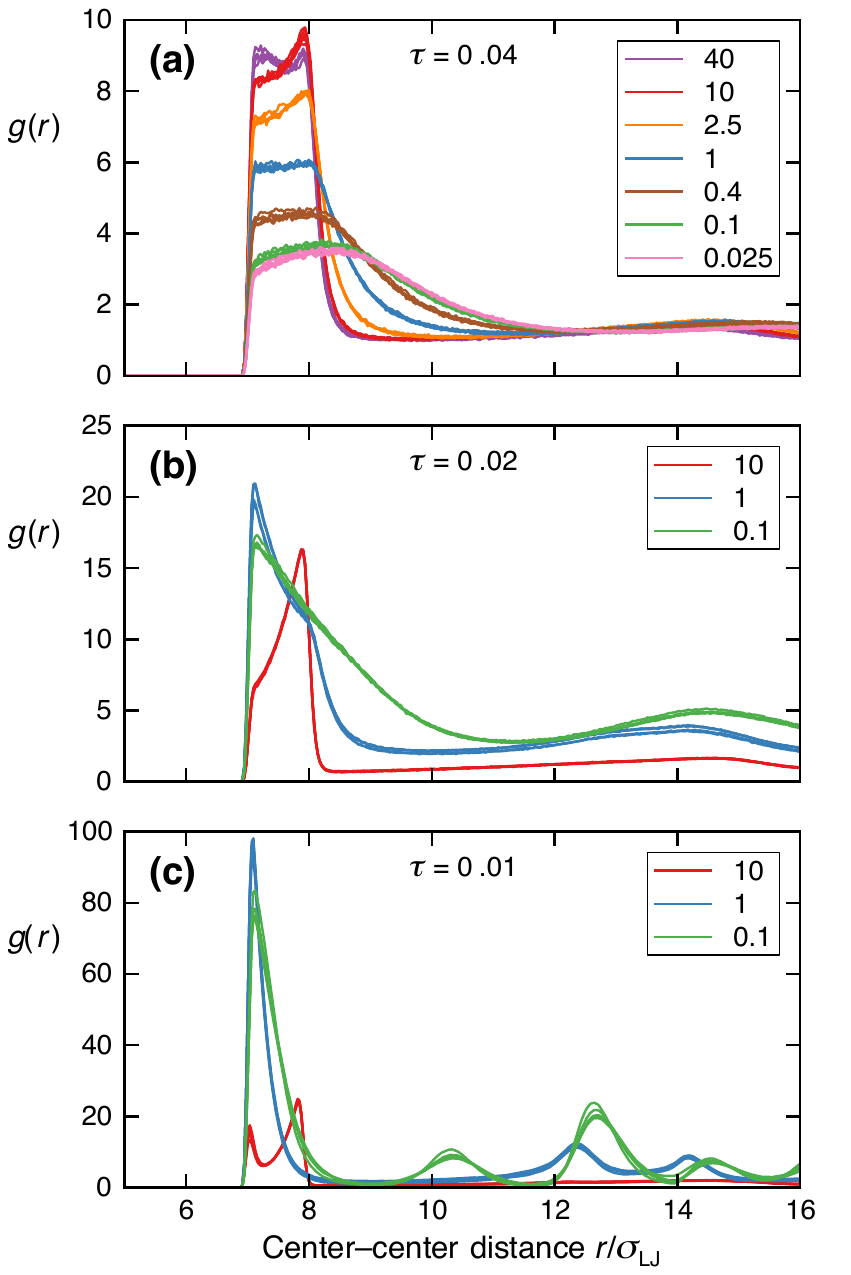}
\caption{Role of dielectric effects in size-asymmetric mixtures of
  charged, polarizable colloids at successively lower reduced
  temperatures (a)~$\tau = 0.04$, (b)~0.02, (c)~0.01.  Each panel shows
  the radial distribution function~$g(r)$ of large colloids for
  different reduced dielectric constants~$\tilde \kappa$.  At high temperatures
  (panel~(a)), the strongest binding occurs for $\tilde \kappa > 1$ as
  polarization charges enhance the large--small binding.  Dielectric
  many-body effects reverse the situation at low temperatures
  (panel~(c)), where $g(r)$ shows the most pronounced structure for
  $\tilde \kappa < 1$. To exclude equilibration artifacts, all runs are
  repeated five times from different initial conditions, with results
  that agree within statistical error.}
\label{fig:correlation} 
\end{figure}

To investigate the properties of this system, we perform large-scale MD
simulations of mixtures containing 100 large colloids and 100 small
colloids.  The excluded-volume interactions between colloids are modeled
via a purely repulsive shifted-truncated Lennard-Jones potential, $4
\epslj [ (\frac{\siglj}{r-\delta})^{12} - (\frac{\siglj}{r-\delta})^6 +
\frac{1}{4}]$ for $r \leqslant 2^{1/6}\siglj + \delta$ with $\delta =
0$, $3 \siglj$, or $6 \siglj$ for small--small, large--small, and
large--large interactions, respectively.
The colloids are placed in a periodic cubic volume, with large-colloid
volume fraction~$5\%$. We take the particles masses to be $m_0$,
yielding a time scale $t_0 = \siglj \sqrt{m_0 / \epslj}$.

Surface bound charges are computed \emph{in each
  time step} using the GMRES algorithm, which converges in 2 or~3
iterations for this system.  The bound charge is discretized using 372
surface patches \emph{per colloid}, placed on a shell of
radius~$3\siglj$, just below the excluded-volume radius~$3.5\siglj$,
resulting in more than $37\,000$ discrete charges in each system.  This
patch density yields a relative error of $\mathcal{O}(10^{-3})$ in the
dielectric interaction energy of a large--small pair at contact.
Starting from a random, nonoverlapping configuration, we investigate
self-assembly by following the system for $1\,000\,000$ time steps of
$0.005t_0$, for a total duration of $5\,000t_0$ per simulation run.  The
first 10\% of each run is discarded.  Temperature is controlled via a
Langevin thermostat with a damping time of~$20t_0$.
To isolate electrostatic effects, we vary the \emph{reduced} temperature
$\tau = k_{\mathrm{B}}T/U_{\mathrm{coul}}$, where $U_{\mathrm{coul}} =
q^2 / (4\pi\varepsilon_0\kappa_{\mathrm{m}} (4\siglj))$ is the Coulomb
interaction of a large and a small colloid at contact.  For simplicity,
we maintain $\epslj = k_{\mathrm{B}}T$.  Then, without loss of
generality, we may reduce the five physical quantities ($q$,
$\kappa_{\mathrm{m}}$, $\kappa_{\mathrm{small}}$,
$\kappa_{\mathrm{large}}$, and $T$) to just two parameters ($\tilde
\kappa=\kappa_{\mathrm{large}}/\kappa_{\mathrm{m}}$ and~$\tau$).

As simulations are performed at successively lower temperatures, the
colloids exhibit a strong tendency to aggregate, as shown by the
large-particle radial distribution function in
Figs.~\ref{fig:correlation}a--c.  First, we consider $g(r)$ for the
highest reduced temperature $\tau = 0.04$
(Fig.~\ref{fig:correlation}a). Here, the colloids are not strongly bound
and $g(r)$ shows a broad peak from $r=7\siglj$ (large particles in
contact, typically bonded by two small colloids) to $r \approx 8\siglj$
(large colloids separated by a small colloid).  Compared to
nonpolarizable colloids ($\tilde \kappa = 1$), colloids with higher
dielectric constant than the solvent ($\tilde \kappa>1$) exhibit a
stronger peak, as the bound-charge interactions become attractive
(Fig.~\ref{fig:sphereenergy}) and the small colloids mediate an
attractive effective interaction between the large ones.  Conversely,
for low-dielectric constant colloids ($\tilde \kappa < 1$) the
polarization charges counteract the Coulombic large--small attraction,
diminishing and broadening the primary peak in~$g(r)$.  The repulsive
interaction between the polarization charges induced by a small colloid
on surrounding colloids amplifies this effect.  Thus, dielectric effects
at this temperature can be qualitatively understood through
decomposition into two-body interactions.

As the reduced temperature is lowered to $\tau = 0.02$
(Fig.~\ref{fig:correlation}b), the situation changes. In the absence of
dielectric effects ($\tilde \kappa=1$) the broad peak observed at $\tau=0.04$
gives way to a prominent contact peak only, signaling the
Coulombic binding of two large colloids by small colloids.  For
$\tilde \kappa=0.1$ the repulsive polarization charges diminish the height of
this peak somewhat.  Most striking, however, is the situation at $\tilde \kappa
= 10$, where a prominent peak at $r=8\siglj$ arises; here
\emph{three-body} interactions qualitatively alter the situation, as the
induced bound charges on the colloids simultaneously enhance the
large--small attraction and yield an effective local repulsion between
the large colloids.

\begin{figure}
\begin{center}
  \setlength\fboxsep{0pt}
  \setlength\fboxrule{1pt}
  \fbox{\includegraphics[scale=0.15]{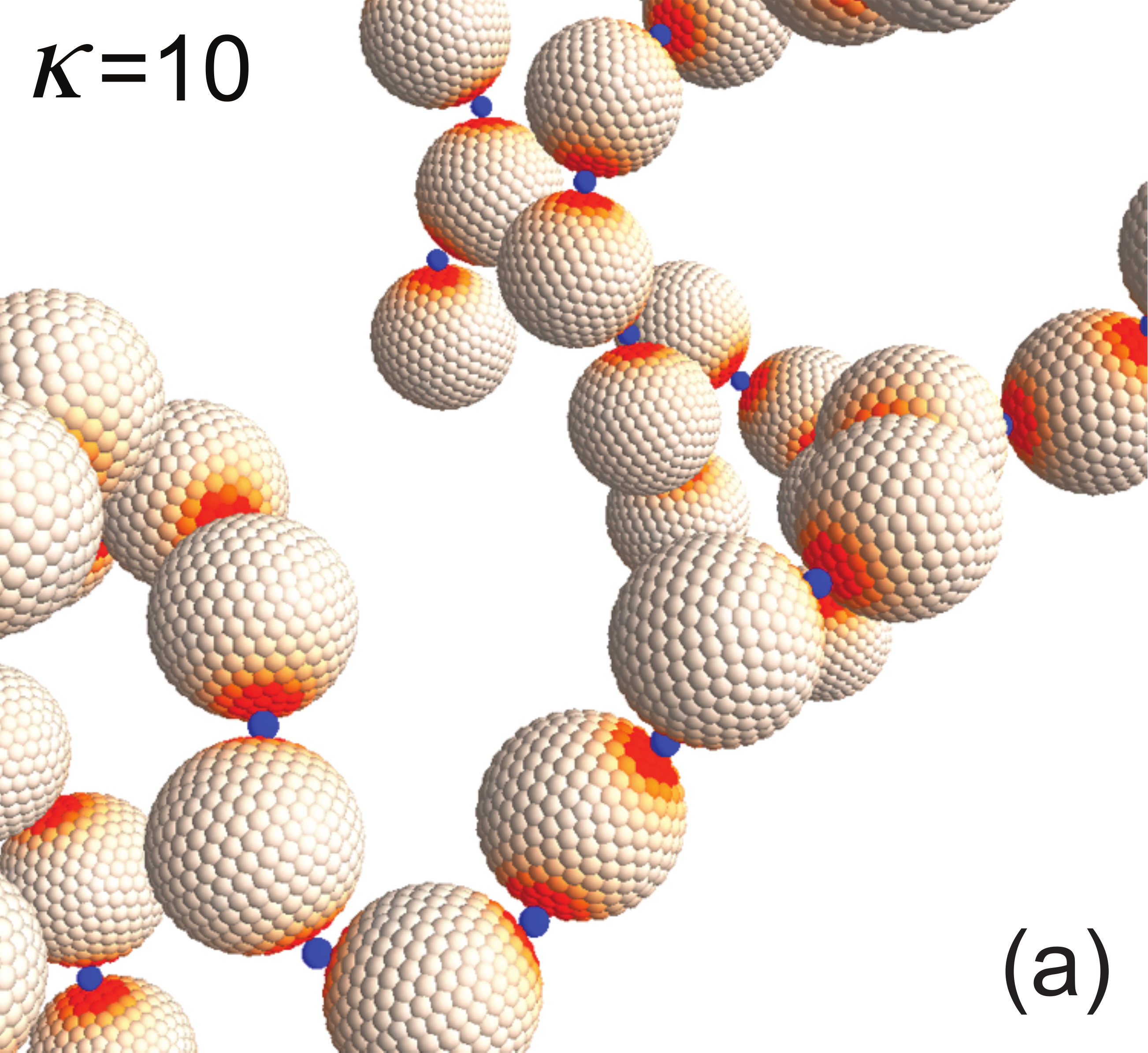}}
  \fbox{\includegraphics[scale=0.15]{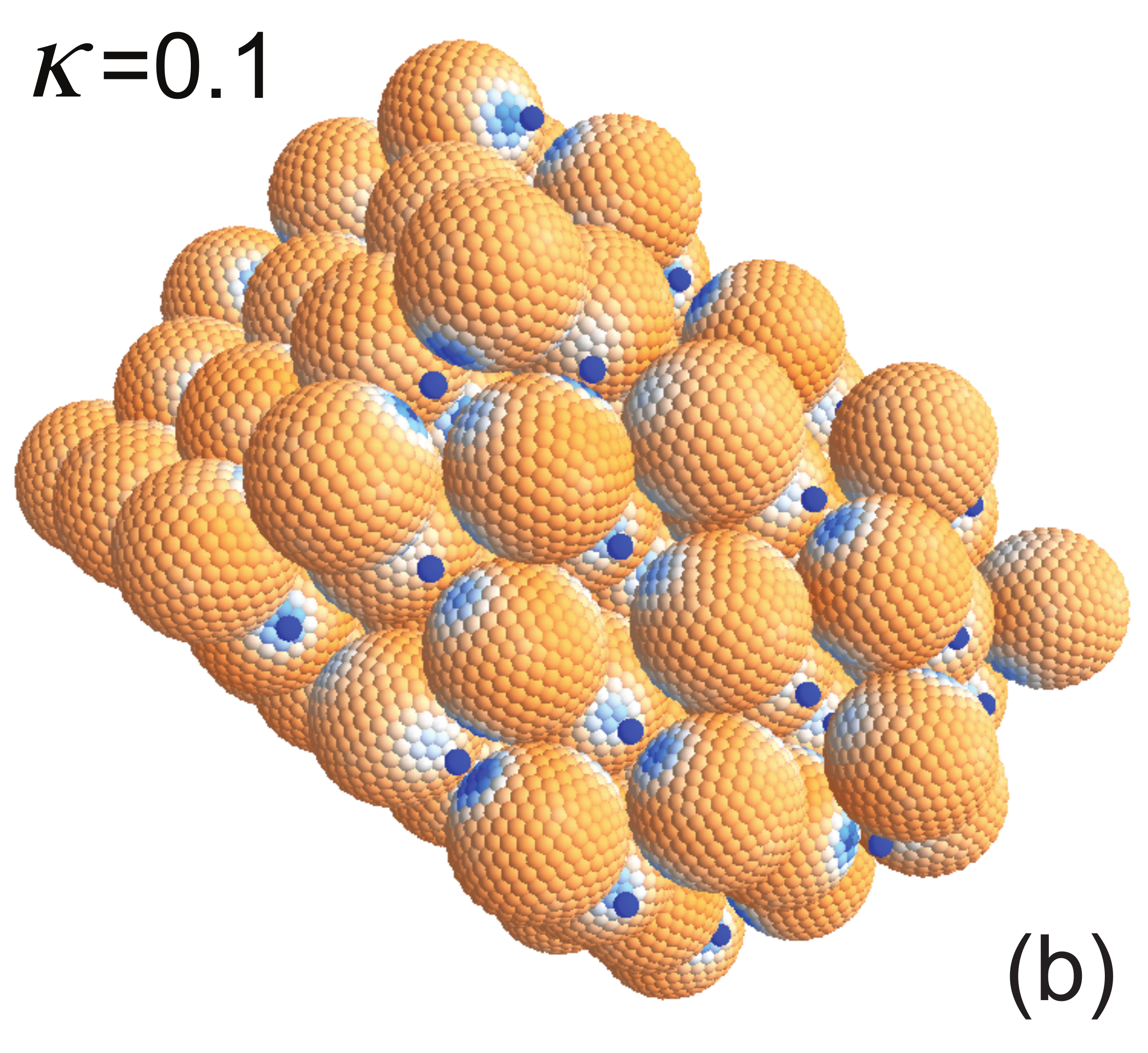}}
  \caption{Example of the importance of polarization charges in
    electrostatic self-assembly.  Low-temperature equilibrium
    configurations of a size-asymmetric binary mixture of colloids at
    reduced dielectric constant (a) $\tilde \kappa = 10$ and (b) $\tilde
    \kappa = 0.1$. The large colloids carry a positive charge and the
    small colloids are negatively charged.  The net surface charge
    density (bare and induced charges) is represented by red (positive)
    and blue (negative) color gradients. The string-like structures in
    panel~(a) arise owing to the prominence of dielectric many-body
    effects (see text). The NaCl structure in panel~(b) is consistent
    with the correlation function in Fig.~\ref{fig:correlation}c.}
\label{fig:configurations}
\end{center}
\end{figure}

Finally, at the lowest reduced temperature $\tau=0.01$
(Fig.~\ref{fig:correlation}c), entropic effects become negligible. The
pair correlation function reveals a complete reversal from the weakly
bound system at $\tau = 0.04$, with the strongest binding and most
ordered structure now occurring at the \emph{lowest} $\tilde \kappa$.
These findings are opposite of the expectations based upon two-body
interactions and result from emergent dielectric many-body interactions.
Indeed, the peaks in $g(r)$ at $\tilde \kappa = 10$, $1$, and $0.1$
correspond to three different structures: strings
(Fig.~\ref{fig:configurations}a), hexagonally-packed ``sheets'' (not
shown), and crystalline aggregates with a sodium chloride structure
(Fig.~\ref{fig:configurations}b), respectively. The string-like
aggregates exhibit a particularly noteworthy example of many-body
effects.  Once two small colloids are bound to diametrically opposite
locations on a large colloid (minimizing their mutual repulsion), the
locally induced (positive) polarization charge in conjunction with the
net-charge requirement Eq.~\eqref{eq:netcharge} results in a negative
polarization charge induced around the ``equator,'' hindering the
association of additional small colloids with this large colloid and
instead promoting the formation of string-like structures.  Not only do
such self-assembled chains offer a striking example of the qualitative
changes that can be induced by polarization effects, but they may also 
provide a (partial) explanation of experimentally observed chain
formation of nanoparticles (for which the many-body effects will be
stronger than for larger colloids) in a range of
solvents~\cite{tang02,liao03,lin05}.

In conclusion, using a newly introduced efficient and generally
applicable method~\cite{barros14a} that permits simulations of a broad
range of systems with fully resolved dielectric many-body effects, we
have explored the role of these effects in the aggregation of colloids
and nanoparticles. We demonstrated that polarization can qualitatively
alter the self-assembled structures. Our approach, which immediately
generalizes to arbitrarily complex dielectric geometries, provides
insight into the underlying mechanisms of recent experimental
observations and makes it possible to exploit dielectric effects to
control colloidal self-assembly.

\begin{acknowledgments}
  This material is based upon work supported by the National Science
  Foundation under Grant Nos.\ DMR-1006430 and DMR-1310211. We thank
  D. Sinkovits for useful discussions and acknowledge allocation of
  computing time on Northwestern University's Quest cluster.
  K.B. acknowledges support from the Theoretical Division and CNLS at
  Los Alamos National Laboratory.
\end{acknowledgments}


\begin{thebibliography}{36}%
\makeatletter
\providecommand \@ifxundefined [1]{%
 \@ifx{#1\undefined}
}%
\providecommand \@ifnum [1]{%
 \ifnum #1\expandafter \@firstoftwo
 \else \expandafter \@secondoftwo
 \fi
}%
\providecommand \@ifx [1]{%
 \ifx #1\expandafter \@firstoftwo
 \else \expandafter \@secondoftwo
 \fi
}%
\providecommand \natexlab [1]{#1}%
\providecommand \enquote  [1]{``#1''}%
\providecommand \bibnamefont  [1]{#1}%
\providecommand \bibfnamefont [1]{#1}%
\providecommand \citenamefont [1]{#1}%
\providecommand \href@noop [0]{\@secondoftwo}%
\providecommand \href [0]{\begingroup \@sanitize@url \@href}%
\providecommand \@href[1]{\@@startlink{#1}\@@href}%
\providecommand \@@href[1]{\endgroup#1\@@endlink}%
\providecommand \@sanitize@url [0]{\catcode `\\12\catcode `\$12\catcode
  `\&12\catcode `\#12\catcode `\^12\catcode `\_12\catcode `\%12\relax}%
\providecommand \@@startlink[1]{}%
\providecommand \@@endlink[0]{}%
\providecommand \url  [0]{\begingroup\@sanitize@url \@url }%
\providecommand \@url [1]{\endgroup\@href {#1}{\urlprefix }}%
\providecommand \urlprefix  [0]{URL }%
\providecommand \Eprint [0]{\href }%
\providecommand \doibase [0]{http://dx.doi.org/}%
\providecommand \selectlanguage [0]{\@gobble}%
\providecommand \bibinfo  [0]{\@secondoftwo}%
\providecommand \bibfield  [0]{\@secondoftwo}%
\providecommand \translation [1]{[#1]}%
\providecommand \BibitemOpen [0]{}%
\providecommand \bibitemStop [0]{}%
\providecommand \bibitemNoStop [0]{.\EOS\space}%
\providecommand \EOS [0]{\spacefactor3000\relax}%
\providecommand \BibitemShut  [1]{\csname bibitem#1\endcsname}%
\let\auto@bib@innerbib\@empty
\bibitem [{\citenamefont {Russel}\ \emph {et~al.}(1989)\citenamefont {Russel},
  \citenamefont {Saville},\ and\ \citenamefont {Schowalter}}]{russel89}%
  \BibitemOpen
  \bibfield  {author} {\bibinfo {author} {\bibfnamefont {W.~B.}\ \bibnamefont
  {Russel}}, \bibinfo {author} {\bibfnamefont {D.~A.}\ \bibnamefont {Saville}},
  \ and\ \bibinfo {author} {\bibfnamefont {W.~R.}\ \bibnamefont {Schowalter}},\
  }\href@noop {} {\emph {\bibinfo {title} {Colloidal Dispersions}}}\ (\bibinfo
  {publisher} {Cambridge University Press},\ \bibinfo {address} {Cambridge,
  U.K.},\ \bibinfo {year} {1989})\BibitemShut {NoStop}%
\bibitem [{\citenamefont {Hunter}(2001)}]{hunter01}%
  \BibitemOpen
  \bibfield  {author} {\bibinfo {author} {\bibfnamefont {R.~J.}\ \bibnamefont
  {Hunter}},\ }\href@noop {} {\emph {\bibinfo {title} {Foundations of {C}olloid
  {S}cience}}},\ \bibinfo {edition} {2nd}\ ed.\ (\bibinfo  {publisher} {Oxford
  University Press},\ \bibinfo {address} {Oxford},\ \bibinfo {year}
  {2001})\BibitemShut {NoStop}%
\bibitem [{\citenamefont {Leunissen}\ \emph {et~al.}(2005)\citenamefont
  {Leunissen}, \citenamefont {Christova}, \citenamefont {Hynninen},
  \citenamefont {Royall}, \citenamefont {Campbell}, \citenamefont {Imhof},
  \citenamefont {Dijkstra}, \citenamefont {van Roij},\ and\ \citenamefont {van
  Blaaderen}}]{leunissen05}%
  \BibitemOpen
  \bibfield  {author} {\bibinfo {author} {\bibfnamefont {M.~E.}\ \bibnamefont
  {Leunissen}}, \bibinfo {author} {\bibfnamefont {C.~G.}\ \bibnamefont
  {Christova}}, \bibinfo {author} {\bibfnamefont {A.-P.}\ \bibnamefont
  {Hynninen}}, \bibinfo {author} {\bibfnamefont {C.~P.}\ \bibnamefont
  {Royall}}, \bibinfo {author} {\bibfnamefont {A.~I.}\ \bibnamefont
  {Campbell}}, \bibinfo {author} {\bibfnamefont {A.}~\bibnamefont {Imhof}},
  \bibinfo {author} {\bibfnamefont {M.}~\bibnamefont {Dijkstra}}, \bibinfo
  {author} {\bibfnamefont {R.}~\bibnamefont {van Roij}}, \ and\ \bibinfo
  {author} {\bibfnamefont {A.}~\bibnamefont {van Blaaderen}},\ }\href@noop {}
  {\bibfield  {journal} {\bibinfo  {journal} {Nature}\ }\textbf {\bibinfo
  {volume} {437}},\ \bibinfo {pages} {235} (\bibinfo {year}
  {2005})}\BibitemShut {NoStop}%
\bibitem [{\citenamefont {Doerr}\ and\ \citenamefont {Yu}(2006)}]{doerr06}%
  \BibitemOpen
  \bibfield  {author} {\bibinfo {author} {\bibfnamefont {T.~P.}\ \bibnamefont
  {Doerr}}\ and\ \bibinfo {author} {\bibfnamefont {Y.-K.}\ \bibnamefont {Yu}},\
  }\href {\doibase 10.1103/PhysRevE.73.061902} {\bibfield  {journal} {\bibinfo
  {journal} {Phys. Rev. E}\ }\textbf {\bibinfo {volume} {73}},\ \bibinfo
  {pages} {061902} (\bibinfo {year} {2006})}\BibitemShut {NoStop}%
\bibitem [{\citenamefont {Levitt}(1978)}]{levitt78}%
  \BibitemOpen
  \bibfield  {author} {\bibinfo {author} {\bibfnamefont {D.~G.}\ \bibnamefont
  {Levitt}},\ }\href@noop {} {\bibfield  {journal} {\bibinfo  {journal}
  {Biophys. J.}\ }\textbf {\bibinfo {volume} {22}},\ \bibinfo {pages} {209}
  (\bibinfo {year} {1978})}\BibitemShut {NoStop}%
\bibitem [{\citenamefont {Hoyles}\ \emph {et~al.}(1998)\citenamefont {Hoyles},
  \citenamefont {Kuyucak},\ and\ \citenamefont {Chung}}]{hoyles98}%
  \BibitemOpen
  \bibfield  {author} {\bibinfo {author} {\bibfnamefont {M.}~\bibnamefont
  {Hoyles}}, \bibinfo {author} {\bibfnamefont {S.}~\bibnamefont {Kuyucak}}, \
  and\ \bibinfo {author} {\bibfnamefont {S.-H.}\ \bibnamefont {Chung}},\
  }\href@noop {} {\bibfield  {journal} {\bibinfo  {journal} {Comp. Phys.
  Comm.}\ }\textbf {\bibinfo {volume} {115}},\ \bibinfo {pages} {45} (\bibinfo
  {year} {1998})}\BibitemShut {NoStop}%
\bibitem [{\citenamefont {Allen}\ \emph {et~al.}(2001)\citenamefont {Allen},
  \citenamefont {Hansen},\ and\ \citenamefont {Melchionna}}]{allen01}%
  \BibitemOpen
  \bibfield  {author} {\bibinfo {author} {\bibfnamefont {R.}~\bibnamefont
  {Allen}}, \bibinfo {author} {\bibfnamefont {J.-P.}\ \bibnamefont {Hansen}}, \
  and\ \bibinfo {author} {\bibfnamefont {S.}~\bibnamefont {Melchionna}},\
  }\href@noop {} {\bibfield  {journal} {\bibinfo  {journal} {Phys. Chem. Chem.
  Phys.}\ }\textbf {\bibinfo {volume} {3}},\ \bibinfo {pages} {4177} (\bibinfo
  {year} {2001})}\BibitemShut {NoStop}%
\bibitem [{\citenamefont {Boda}\ \emph {et~al.}(2004)\citenamefont {Boda},
  \citenamefont {Gillespie}, \citenamefont {Nonner}, \citenamefont
  {Henderson},\ and\ \citenamefont {Eisenberg}}]{boda04}%
  \BibitemOpen
  \bibfield  {author} {\bibinfo {author} {\bibfnamefont {D.}~\bibnamefont
  {Boda}}, \bibinfo {author} {\bibfnamefont {D.}~\bibnamefont {Gillespie}},
  \bibinfo {author} {\bibfnamefont {W.}~\bibnamefont {Nonner}}, \bibinfo
  {author} {\bibfnamefont {D.}~\bibnamefont {Henderson}}, \ and\ \bibinfo
  {author} {\bibfnamefont {B.}~\bibnamefont {Eisenberg}},\ }\href@noop {}
  {\bibfield  {journal} {\bibinfo  {journal} {Phys. Rev. E}\ }\textbf {\bibinfo
  {volume} {69}},\ \bibinfo {pages} {046702} (\bibinfo {year}
  {2004})}\BibitemShut {NoStop}%
\bibitem [{\citenamefont {Tyagi}\ \emph {et~al.}(2010)\citenamefont {Tyagi},
  \citenamefont {S{\"{u}}zen}, \citenamefont {Sega}, \citenamefont {Barbosa},
  \citenamefont {Kantorovich},\ and\ \citenamefont {Holm}}]{tyagi10}%
  \BibitemOpen
  \bibfield  {author} {\bibinfo {author} {\bibfnamefont {S.}~\bibnamefont
  {Tyagi}}, \bibinfo {author} {\bibfnamefont {M.}~\bibnamefont {S{\"{u}}zen}},
  \bibinfo {author} {\bibfnamefont {M.}~\bibnamefont {Sega}}, \bibinfo {author}
  {\bibfnamefont {M.}~\bibnamefont {Barbosa}}, \bibinfo {author} {\bibfnamefont
  {S.~S.}\ \bibnamefont {Kantorovich}}, \ and\ \bibinfo {author} {\bibfnamefont
  {C.}~\bibnamefont {Holm}},\ }\href@noop {} {\bibfield  {journal} {\bibinfo
  {journal} {J. Chem. Phys.}\ }\textbf {\bibinfo {volume} {132}},\ \bibinfo
  {pages} {154112} (\bibinfo {year} {2010})}\BibitemShut {NoStop}%
\bibitem [{\citenamefont {Jadhao}\ \emph {et~al.}(2012)\citenamefont {Jadhao},
  \citenamefont {Solis},\ and\ \citenamefont {Olvera de~la Cruz}}]{jadhao12}%
  \BibitemOpen
  \bibfield  {author} {\bibinfo {author} {\bibfnamefont {V.}~\bibnamefont
  {Jadhao}}, \bibinfo {author} {\bibfnamefont {F.~J.}\ \bibnamefont {Solis}}, \
  and\ \bibinfo {author} {\bibfnamefont {M.}~\bibnamefont {Olvera de~la
  Cruz}},\ }\href {\doibase 10.1103/PhysRevLett.109.223905} {\bibfield
  {journal} {\bibinfo  {journal} {Phys. Rev. Lett.}\ }\textbf {\bibinfo
  {volume} {109}},\ \bibinfo {pages} {223905} (\bibinfo {year}
  {2012})}\BibitemShut {NoStop}%
\bibitem [{\citenamefont {Croxton}\ \emph {et~al.}(1981)\citenamefont
  {Croxton}, \citenamefont {McQuarrie}, \citenamefont {Patey}, \citenamefont
  {Torrie},\ and\ \citenamefont {Valleau}}]{croxton81}%
  \BibitemOpen
  \bibfield  {author} {\bibinfo {author} {\bibfnamefont {T.}~\bibnamefont
  {Croxton}}, \bibinfo {author} {\bibfnamefont {D.~A.}\ \bibnamefont
  {McQuarrie}}, \bibinfo {author} {\bibfnamefont {G.~N.}\ \bibnamefont
  {Patey}}, \bibinfo {author} {\bibfnamefont {G.~M.}\ \bibnamefont {Torrie}}, \
  and\ \bibinfo {author} {\bibfnamefont {J.~P.}\ \bibnamefont {Valleau}},\
  }\href {\doibase 10.1139/v81-295} {\bibfield  {journal} {\bibinfo  {journal}
  {Can. J. Chem.}\ }\textbf {\bibinfo {volume} {59}},\ \bibinfo {pages} {1998}
  (\bibinfo {year} {1981})}\BibitemShut {NoStop}%
\bibitem [{\citenamefont {Torrie}\ \emph {et~al.}(1982)\citenamefont {Torrie},
  \citenamefont {Valleau},\ and\ \citenamefont {Patey}}]{torrie82}%
  \BibitemOpen
  \bibfield  {author} {\bibinfo {author} {\bibfnamefont {G.~M.}\ \bibnamefont
  {Torrie}}, \bibinfo {author} {\bibfnamefont {J.~P.}\ \bibnamefont {Valleau}},
  \ and\ \bibinfo {author} {\bibfnamefont {G.~N.}\ \bibnamefont {Patey}},\
  }\href {\doibase 10.1063/1.443541} {\bibfield  {journal} {\bibinfo  {journal}
  {J. Chem. Phys.}\ }\textbf {\bibinfo {volume} {76}},\ \bibinfo {pages} {4615}
  (\bibinfo {year} {1982})}\BibitemShut {NoStop}%
\bibitem [{\citenamefont {Kjellander}\ and\ \citenamefont
  {Mar\v{c}elja}(1985)}]{kjellander85}%
  \BibitemOpen
  \bibfield  {author} {\bibinfo {author} {\bibfnamefont {R.}~\bibnamefont
  {Kjellander}}\ and\ \bibinfo {author} {\bibfnamefont {S.}~\bibnamefont
  {Mar\v{c}elja}},\ }\href {\doibase 10.1063/1.448350} {\bibfield  {journal}
  {\bibinfo  {journal} {J. Chem. Phys.}\ }\textbf {\bibinfo {volume} {82}},\
  \bibinfo {pages} {2122} (\bibinfo {year} {1985})}\BibitemShut {NoStop}%
\bibitem [{\citenamefont {Messina}(2002)}]{messina02a}%
  \BibitemOpen
  \bibfield  {author} {\bibinfo {author} {\bibfnamefont {R.}~\bibnamefont
  {Messina}},\ }\href@noop {} {\bibfield  {journal} {\bibinfo  {journal} {J.
  Chem. Phys.}\ }\textbf {\bibinfo {volume} {117}},\ \bibinfo {pages} {11062}
  (\bibinfo {year} {2002})}\BibitemShut {NoStop}%
\bibitem [{\citenamefont {Re\v{s}\v{c}i\v{c}}\ and\ \citenamefont
  {Linse}(2008)}]{rescic08}%
  \BibitemOpen
  \bibfield  {author} {\bibinfo {author} {\bibfnamefont {J.}~\bibnamefont
  {Re\v{s}\v{c}i\v{c}}}\ and\ \bibinfo {author} {\bibfnamefont
  {P.}~\bibnamefont {Linse}},\ }\href {\doibase 10.1063/1.2971038} {\bibfield
  {journal} {\bibinfo  {journal} {J. Chem. Phys.}\ }\textbf {\bibinfo {volume}
  {129}},\ \bibinfo {eid} {114505} (\bibinfo {year} {2008})}\BibitemShut
  {NoStop}%
\bibitem [{\citenamefont {Barros}\ \emph {et~al.}(2014)\citenamefont {Barros},
  \citenamefont {Sinkovits},\ and\ \citenamefont {Luijten}}]{barros14a}%
  \BibitemOpen
  \bibfield  {author} {\bibinfo {author} {\bibfnamefont {K.}~\bibnamefont
  {Barros}}, \bibinfo {author} {\bibfnamefont {D.}~\bibnamefont {Sinkovits}}, \
  and\ \bibinfo {author} {\bibfnamefont {E.}~\bibnamefont {Luijten}},\
  }\href@noop {} {\bibfield  {journal} {\bibinfo  {journal} {J. Chem. Phys.}\
  }\textbf {\bibinfo {volume} {140}},\ \bibinfo {pages} {064903} (\bibinfo
  {year} {2014})}\BibitemShut {NoStop}%
\bibitem [{\citenamefont {Jackson}(1999)}]{jackson99}%
  \BibitemOpen
  \bibfield  {author} {\bibinfo {author} {\bibfnamefont {J.~D.}\ \bibnamefont
  {Jackson}},\ }\href@noop {} {\emph {\bibinfo {title} {Classical
  Electrodynamics}}},\ \bibinfo {edition} {3rd}\ ed.\ (\bibinfo  {publisher}
  {Wiley},\ \bibinfo {address} {New York},\ \bibinfo {year} {1999})\BibitemShut
  {NoStop}%
\bibitem [{\citenamefont {Israelachvili}(2011)}]{israel2011}%
  \BibitemOpen
  \bibfield  {author} {\bibinfo {author} {\bibfnamefont {J.~N.}\ \bibnamefont
  {Israelachvili}},\ }\href@noop {} {\emph {\bibinfo {title} {Intermolecular
  and Surface Forces}}},\ \bibinfo {edition} {3rd}\ ed.\ (\bibinfo  {publisher}
  {Academic},\ \bibinfo {address} {San Diego},\ \bibinfo {year}
  {2011})\BibitemShut {NoStop}%
\bibitem [{\citenamefont {Hoshi}\ \emph {et~al.}(1987)\citenamefont {Hoshi},
  \citenamefont {Sakurai}, \citenamefont {Inoue},\ and\ \citenamefont
  {Ch{\^u}j{\^o}}}]{hoshi87}%
  \BibitemOpen
  \bibfield  {author} {\bibinfo {author} {\bibfnamefont {H.}~\bibnamefont
  {Hoshi}}, \bibinfo {author} {\bibfnamefont {M.}~\bibnamefont {Sakurai}},
  \bibinfo {author} {\bibfnamefont {Y.}~\bibnamefont {Inoue}}, \ and\ \bibinfo
  {author} {\bibfnamefont {R.}~\bibnamefont {Ch{\^u}j{\^o}}},\ }\href@noop {}
  {\bibfield  {journal} {\bibinfo  {journal} {J. Chem. Phys.}\ }\textbf
  {\bibinfo {volume} {87}},\ \bibinfo {pages} {1107} (\bibinfo {year}
  {1987})}\BibitemShut {NoStop}%
\bibitem [{\citenamefont {Saad}\ and\ \citenamefont {Schultz}(1986)}]{saad86}%
  \BibitemOpen
  \bibfield  {author} {\bibinfo {author} {\bibfnamefont {Y.}~\bibnamefont
  {Saad}}\ and\ \bibinfo {author} {\bibfnamefont {M.~H.}\ \bibnamefont
  {Schultz}},\ }\href {\doibase 10.1137/0907058} {\bibfield  {journal}
  {\bibinfo  {journal} {SIAM J. Sci. Stat. Comput.}\ }\textbf {\bibinfo
  {volume} {7}},\ \bibinfo {pages} {856} (\bibinfo {year} {1986})}\BibitemShut
  {NoStop}%
\bibitem [{\citenamefont {van~der Vorst}(1992)}]{vandervorst92}%
  \BibitemOpen
  \bibfield  {author} {\bibinfo {author} {\bibfnamefont {H.}~\bibnamefont
  {van~der Vorst}},\ }\href {\doibase 10.1137/0913035} {\bibfield  {journal}
  {\bibinfo  {journal} {SIAM J. Sci. Stat. Comput.}\ }\textbf {\bibinfo
  {volume} {13}},\ \bibinfo {pages} {631} (\bibinfo {year} {1992})}\BibitemShut
  {NoStop}%
\bibitem [{\citenamefont {Greengard}\ and\ \citenamefont
  {Rokhlin}(1997)}]{greengard97}%
  \BibitemOpen
  \bibfield  {author} {\bibinfo {author} {\bibfnamefont {L.}~\bibnamefont
  {Greengard}}\ and\ \bibinfo {author} {\bibfnamefont {V.}~\bibnamefont
  {Rokhlin}},\ }\href@noop {} {\bibfield  {journal} {\bibinfo  {journal} {Acta
  Numerica}\ }\textbf {\bibinfo {volume} {6}},\ \bibinfo {pages} {229}
  (\bibinfo {year} {1997})}\BibitemShut {NoStop}%
\bibitem [{\citenamefont {Sagui}\ and\ \citenamefont {Darden}(2001)}]{sagui01}%
  \BibitemOpen
  \bibfield  {author} {\bibinfo {author} {\bibfnamefont {C.}~\bibnamefont
  {Sagui}}\ and\ \bibinfo {author} {\bibfnamefont {T.}~\bibnamefont {Darden}},\
  }\href {\doibase 10.1063/1.1352646} {\bibfield  {journal} {\bibinfo
  {journal} {J. Chem. Phys.}\ }\textbf {\bibinfo {volume} {114}},\ \bibinfo
  {pages} {6578} (\bibinfo {year} {2001})}\BibitemShut {NoStop}%
\bibitem [{\citenamefont {Pohl}(1951)}]{pohl51}%
  \BibitemOpen
  \bibfield  {author} {\bibinfo {author} {\bibfnamefont {H.~A.}\ \bibnamefont
  {Pohl}},\ }\href@noop {} {\bibfield  {journal} {\bibinfo  {journal} {J. Appl.
  Phys.}\ }\textbf {\bibinfo {volume} {22}},\ \bibinfo {pages} {869} (\bibinfo
  {year} {1951})}\BibitemShut {NoStop}%
\bibitem [{\citenamefont {Jones}(1995)}]{jones95}%
  \BibitemOpen
  \bibfield  {author} {\bibinfo {author} {\bibfnamefont {T.~B.}\ \bibnamefont
  {Jones}},\ }\href@noop {} {\emph {\bibinfo {title} {Electromechanics of
  particles}}}\ (\bibinfo  {publisher} {Cambridge University Press},\ \bibinfo
  {address} {Cambridge, U.K.},\ \bibinfo {year} {1995})\BibitemShut {NoStop}%
\bibitem [{\citenamefont {Romero-Enrique}\ \emph {et~al.}(2000)\citenamefont
  {Romero-Enrique}, \citenamefont {Orkoulas}, \citenamefont {Panagiotopoulos},\
  and\ \citenamefont {Fisher}}]{romero-enrique00}%
  \BibitemOpen
  \bibfield  {author} {\bibinfo {author} {\bibfnamefont {J.~M.}\ \bibnamefont
  {Romero-Enrique}}, \bibinfo {author} {\bibfnamefont {G.}~\bibnamefont
  {Orkoulas}}, \bibinfo {author} {\bibfnamefont {A.~Z.}\ \bibnamefont
  {Panagiotopoulos}}, \ and\ \bibinfo {author} {\bibfnamefont {M.~E.}\
  \bibnamefont {Fisher}},\ }\href@noop {} {\bibfield  {journal} {\bibinfo
  {journal} {Phys. Rev. Lett.}\ }\textbf {\bibinfo {volume} {85}},\ \bibinfo
  {pages} {4558} (\bibinfo {year} {2000})}\BibitemShut {NoStop}%
\bibitem [{\citenamefont {Liu}\ and\ \citenamefont {Luijten}(2004)}]{liu04b}%
  \BibitemOpen
  \bibfield  {author} {\bibinfo {author} {\bibfnamefont {J.}~\bibnamefont
  {Liu}}\ and\ \bibinfo {author} {\bibfnamefont {E.}~\bibnamefont {Luijten}},\
  }\href@noop {} {\bibfield  {journal} {\bibinfo  {journal} {Phys. Rev. Lett.}\
  }\textbf {\bibinfo {volume} {93}},\ \bibinfo {pages} {247802} (\bibinfo
  {year} {2004})}\BibitemShut {NoStop}%
\bibitem [{\citenamefont {Van~Tomme}\ \emph {et~al.}(2008)\citenamefont
  {Van~Tomme}, \citenamefont {van Nostrum}, \citenamefont {Dijkstra},
  \citenamefont {Smedt},\ and\ \citenamefont {Hennink}}]{vantomme08}%
  \BibitemOpen
  \bibfield  {author} {\bibinfo {author} {\bibfnamefont {S.~R.}\ \bibnamefont
  {Van~Tomme}}, \bibinfo {author} {\bibfnamefont {C.~F.}\ \bibnamefont {van
  Nostrum}}, \bibinfo {author} {\bibfnamefont {M.}~\bibnamefont {Dijkstra}},
  \bibinfo {author} {\bibfnamefont {S.~C.~D.}\ \bibnamefont {Smedt}}, \ and\
  \bibinfo {author} {\bibfnamefont {W.~E.}\ \bibnamefont {Hennink}},\ }\href
  {\doibase 10.1016/j.ejpb.2008.05.013} {\bibfield  {journal} {\bibinfo
  {journal} {Eur. J. Pharm. Biopharm.}\ }\textbf {\bibinfo {volume} {70}},\
  \bibinfo {pages} {522} (\bibinfo {year} {2008})}\BibitemShut {NoStop}%
\bibitem [{\citenamefont {Kim}\ \emph {et~al.}(2006)\citenamefont {Kim},
  \citenamefont {Sofo}, \citenamefont {Velegol}, \citenamefont {Cole},\ and\
  \citenamefont {Lucas}}]{kim06}%
  \BibitemOpen
  \bibfield  {author} {\bibinfo {author} {\bibfnamefont {H.-Y.}\ \bibnamefont
  {Kim}}, \bibinfo {author} {\bibfnamefont {J.~O.}\ \bibnamefont {Sofo}},
  \bibinfo {author} {\bibfnamefont {D.}~\bibnamefont {Velegol}}, \bibinfo
  {author} {\bibfnamefont {M.~W.}\ \bibnamefont {Cole}}, \ and\ \bibinfo
  {author} {\bibfnamefont {A.~A.}\ \bibnamefont {Lucas}},\ }\href@noop {}
  {\bibfield  {journal} {\bibinfo  {journal} {J. Chem. Phys.}\ }\textbf
  {\bibinfo {volume} {124}},\ \bibinfo {pages} {074504} (\bibinfo {year}
  {2006})}\BibitemShut {NoStop}%
\bibitem [{\citenamefont {Cole}\ \emph {et~al.}(2010)\citenamefont {Cole},
  \citenamefont {Gergidis}, \citenamefont {McNutt}, \citenamefont {Velegol},
  \citenamefont {Kim},\ and\ \citenamefont {Bond}}]{cole10}%
  \BibitemOpen
  \bibfield  {author} {\bibinfo {author} {\bibfnamefont {M.~W.}\ \bibnamefont
  {Cole}}, \bibinfo {author} {\bibfnamefont {L.~N.}\ \bibnamefont {Gergidis}},
  \bibinfo {author} {\bibfnamefont {J.~P.}\ \bibnamefont {McNutt}}, \bibinfo
  {author} {\bibfnamefont {D.}~\bibnamefont {Velegol}}, \bibinfo {author}
  {\bibfnamefont {H.-Y.}\ \bibnamefont {Kim}}, \ and\ \bibinfo {author}
  {\bibfnamefont {Z.~K.}\ \bibnamefont {Bond}},\ }\href@noop {} {\bibfield
  {journal} {\bibinfo  {journal} {J. Nanophotonics}\ }\textbf {\bibinfo
  {volume} {4}},\ \bibinfo {pages} {041560} (\bibinfo {year}
  {2010})}\BibitemShut {NoStop}%
\bibitem [{\citenamefont {Ninham}\ and\ \citenamefont
  {Parsegian}(1971)}]{ninham71}%
  \BibitemOpen
  \bibfield  {author} {\bibinfo {author} {\bibfnamefont {B.~W.}\ \bibnamefont
  {Ninham}}\ and\ \bibinfo {author} {\bibfnamefont {V.~A.}\ \bibnamefont
  {Parsegian}},\ }\href@noop {} {\bibfield  {journal} {\bibinfo  {journal} {J.
  Theor. Biol.}\ }\textbf {\bibinfo {volume} {31}},\ \bibinfo {pages} {405}
  (\bibinfo {year} {1971})}\BibitemShut {NoStop}%
\bibitem [{\citenamefont {Chan}\ and\ \citenamefont {Mitchell}(1983)}]{chan83}%
  \BibitemOpen
  \bibfield  {author} {\bibinfo {author} {\bibfnamefont {D.~Y.~C.}\
  \bibnamefont {Chan}}\ and\ \bibinfo {author} {\bibfnamefont {D.~J.}\
  \bibnamefont {Mitchell}},\ }\href@noop {} {\bibfield  {journal} {\bibinfo
  {journal} {J. Colloid Interface Sci.}\ }\textbf {\bibinfo {volume} {95}},\
  \bibinfo {pages} {193} (\bibinfo {year} {1983})}\BibitemShut {NoStop}%
\bibitem [{\citenamefont {Popa}\ \emph {et~al.}(2010)\citenamefont {Popa},
  \citenamefont {Sinha}, \citenamefont {Finessi}, \citenamefont {Maroni},
  \citenamefont {Papastavrou},\ and\ \citenamefont {Borkovec}}]{popa10}%
  \BibitemOpen
  \bibfield  {author} {\bibinfo {author} {\bibfnamefont {I.}~\bibnamefont
  {Popa}}, \bibinfo {author} {\bibfnamefont {P.}~\bibnamefont {Sinha}},
  \bibinfo {author} {\bibfnamefont {M.}~\bibnamefont {Finessi}}, \bibinfo
  {author} {\bibfnamefont {P.}~\bibnamefont {Maroni}}, \bibinfo {author}
  {\bibfnamefont {G.}~\bibnamefont {Papastavrou}}, \ and\ \bibinfo {author}
  {\bibfnamefont {M.}~\bibnamefont {Borkovec}},\ }\href@noop {} {\bibfield
  {journal} {\bibinfo  {journal} {Phys. Rev. Lett.}\ }\textbf {\bibinfo
  {volume} {104}},\ \bibinfo {pages} {228301} (\bibinfo {year}
  {2010})}\BibitemShut {NoStop}%
\bibitem [{\citenamefont {Tang}\ \emph {et~al.}(2002)\citenamefont {Tang},
  \citenamefont {Kotov},\ and\ \citenamefont {Giersig}}]{tang02}%
  \BibitemOpen
  \bibfield  {author} {\bibinfo {author} {\bibfnamefont {Z.}~\bibnamefont
  {Tang}}, \bibinfo {author} {\bibfnamefont {N.~A.}\ \bibnamefont {Kotov}}, \
  and\ \bibinfo {author} {\bibfnamefont {M.}~\bibnamefont {Giersig}},\
  }\href@noop {} {\bibfield  {journal} {\bibinfo  {journal} {Science}\ }\textbf
  {\bibinfo {volume} {297}},\ \bibinfo {pages} {237} (\bibinfo {year}
  {2002})}\BibitemShut {NoStop}%
\bibitem [{\citenamefont {Liao}\ \emph {et~al.}(2003)\citenamefont {Liao},
  \citenamefont {Chen}, \citenamefont {Xu}, \citenamefont {Ge}, \citenamefont
  {Wang}, \citenamefont {Huang},\ and\ \citenamefont {Gu}}]{liao03}%
  \BibitemOpen
  \bibfield  {author} {\bibinfo {author} {\bibfnamefont {J.~H.}\ \bibnamefont
  {Liao}}, \bibinfo {author} {\bibfnamefont {K.~J.}\ \bibnamefont {Chen}},
  \bibinfo {author} {\bibfnamefont {L.~N.}\ \bibnamefont {Xu}}, \bibinfo
  {author} {\bibfnamefont {C.~W.}\ \bibnamefont {Ge}}, \bibinfo {author}
  {\bibfnamefont {J.}~\bibnamefont {Wang}}, \bibinfo {author} {\bibfnamefont
  {L.}~\bibnamefont {Huang}}, \ and\ \bibinfo {author} {\bibfnamefont
  {N.}~\bibnamefont {Gu}},\ }\href@noop {} {\bibfield  {journal} {\bibinfo
  {journal} {Appl. Phys. A}\ }\textbf {\bibinfo {volume} {76}},\ \bibinfo
  {pages} {541} (\bibinfo {year} {2003})}\BibitemShut {NoStop}%
\bibitem [{\citenamefont {Lin}\ \emph {et~al.}(2005)\citenamefont {Lin},
  \citenamefont {Li}, \citenamefont {Dujardin}, \citenamefont {Girard},\ and\
  \citenamefont {Mann}}]{lin05}%
  \BibitemOpen
  \bibfield  {author} {\bibinfo {author} {\bibfnamefont {S.}~\bibnamefont
  {Lin}}, \bibinfo {author} {\bibfnamefont {M.}~\bibnamefont {Li}}, \bibinfo
  {author} {\bibfnamefont {E.}~\bibnamefont {Dujardin}}, \bibinfo {author}
  {\bibfnamefont {C.}~\bibnamefont {Girard}}, \ and\ \bibinfo {author}
  {\bibfnamefont {S.}~\bibnamefont {Mann}},\ }\href@noop {} {\bibfield
  {journal} {\bibinfo  {journal} {Adv. Mater.}\ }\textbf {\bibinfo {volume}
  {17}},\ \bibinfo {pages} {2553} (\bibinfo {year} {2005})}\BibitemShut
  {NoStop}%
\end{thebibliography}
%

\end{document}